\documentclass[twocolumn,aps,pra,showpacs]{revtex4}
\usepackage{amsmath}

\begin{document}

\title{Comment on ``Quantum nonlocality for a three-particle nonmaximally entangled state without inequalities''}
\author{Jos\'{e} L. Cereceda}
\email{jl.cereceda@telefonica.net}
\affiliation{C/Alto del Le\'{o}n 8, 4A, 28038 Madrid, Spain}
\date{\today}


\begin{abstract}
In a recent Brief Report, Zheng [S-B. Zheng, Phys. Rev. A \textbf{66}, 014103 (2002)] has given a proof of nonlocality without using inequalities for three spin-1/2 particles in the nonmaximally entangled state $|\psi_{\text{1,2,3}}\rangle = \cos\theta|+_1\rangle|+_2\rangle|+_3\rangle+ i\sin\theta|-_1\rangle|-_2\rangle|-_3\rangle$. Here we show that Zheng's proof is not correct. Indeed it is the case that, for the experiment considered by Zheng, the only state that admits a nonlocality proof without inequalities is the maximally entangled state.
\end{abstract}



\maketitle

In Ref.~\cite{Zheng}, Zheng attempted to demonstrate quantum nonlocality without using inequality for three spacelike separated spin-1/2 particles in the nonmaximally entangled state
\begin{equation}
|\psi_{\text{1,2,3}}\rangle = \cos\theta|+_1\rangle|+_2\rangle|+_3\rangle+ i\sin\theta|-_1\rangle|-_2\rangle|-_3\rangle,
\end{equation}
with $0<\theta<\pi/4$, and where $|+_i\rangle$ and $|-_i\rangle$ denote the spin-up and -down states along the $z$ axis for the $i$th particle ($i=1,2,3$). Zheng's argument for nonlocality is as follows. Consider the physical, two-valued observables $E_i$ and $U_i$ with corresponding operators $\hat{E}_i$ and $\hat{U}_i$ appearing, respectively, in Eqs.~(6) and (7) of the paper \cite{Zheng}. The result of any measurement is labeled $\pm 1$. The relevant quantum-mechanical predictions for the state (1) are
\begin{align}
& \text{If}\,\, E_1 =1,\,\,\text{then}\,\, U_2 U_3 =-1,  \\
& \text{If}\,\, E_2 =1,\,\,\text{then}\,\, U_1 U_3 =-1,  \\
& \text{If}\,\, E_3 =1,\,\,\text{then}\,\, U_1 U_2 =-1.
\end{align}
In addition to this, the probability of obtaining the result $E_1=E_2=E_3=1$ is not zero for the state (1). So consider a run of measurements for which $E_1=E_2=E_3=1$ is obtained. Now, from Eq.~(2), and assuming local hidden variable theory, it follows that if $U_2$ and $U_3$ had been measured in the same run of measurements one should have obtained $U_2U_3=-1$. Applying the EPR reality criterion \cite{EPR} to the system formed by particles 2 and 3, one can thus assign an element of physical reality to the \textit{product\/} observable $U_2U_3$ with value $-1$. Similarly, from Eqs.~(3) and (4), and according to local hidden variable theory, there must exist elements of reality for the product observables $U_1U_3$ and $U_1U_2$, each having the value $-1$. Thus the product of the values of these three elements of reality is
\begin{equation}
(U_2U_3)(U_1U_3)(U_1U_2) = -1.
\end{equation}
On the other hand, if one assumes that elements of reality exist for the individual observables $U_i$, each having the value $1$ or $-1$, then
\begin{equation}
(U_2U_3)(U_1U_3)(U_1U_2) = 1,
\end{equation}
in contradiction with Eq.~(5). From this, Zheng deduces the ''inconsistency hidden in the local hidden variable theory''.

This conclusion, however, is not justified in the present experimental situation since no element of physical reality can be assigned to the individual observables $U_i$ on the basis of the EPR criterion. So, for example, the fact that $E_1=1$, does not provide us with any information about the values of $U_2$ and $U_3$ separately, but only about the value of the product observable $U_2U_3$ as a whole (see Eq.~(2)). Indeed, there is maximal uncertainty regarding the measurement values of $U_2$ and $U_3$ when $E_1=1$ is obtained. Therefore the EPR reality criterion cannot be invoked to assert the existence of elements of reality for any of the $U_i$'s individually, and then there is no contradiction between the local hidden variable theory and the relevant quantum-mechanical predictions for the experiment in question. It is only when one introduces the \textit{ad hoc\/} assumption that elements of reality do exist for the individual $U_i$'s that a contradiction arises.

In order to reveal the nonlocality of the state (1) it is necessary to use a Bell-type inequality and the \textit{statistical\/} predictions of quantum mechanics. The appropriate Bell inequality for Zheng's experiment is the one derived by Mermin for $N$ spin-1/2 particles and two alternative dichotomic measurements per particle \cite{Mermin90a}. For three spin-1/2 particles, Mermin's inequality reads
\begin{align}
|\mathcal{M}| \equiv | & \langle E_1E_2E_3\rangle - \langle E_1U_2U_3\rangle
  - \langle U_1E_2U_3\rangle  \nonumber \\
 & - \langle U_1U_2E_3\rangle| \leq 2.
\end{align}
For the state (1), quantum mechanics predicts
\begin{equation}
\mathcal{M}_{\text{QM}} = \sin^4 2\theta - \cos^4 2\theta + 3\sin^2 2\theta.
\end{equation}
Within the range $0<\theta<\pi/4$, inequality (7) is violated by the quantum prediction (8) provided that $\theta > (\arccos\sqrt{2/5})/2 \approx 25.38^{\circ}$. It is important to note that this result is consistent with the fact \cite{SG,ZBLW} according to which Mermin's inequality (7) \textit{cannot\/} be violated by those states (1) for which $\theta \leq 15^{\circ}$. On the other hand, the maximal violation of the inequality (7) occurs for the maximally entangled state, $\theta=\pi/4$. For this case we have the perfect correlations, $\langle E_1E_2E_3\rangle = - \langle E_1U_2U_3\rangle = - \langle U_1E_2U_3\rangle = - \langle U_1U_2E_3\rangle =\nolinebreak 1$.

We thus conclude that, for the experimental setup proposed in Ref.~\cite{Zheng}, the class of states in Eq.~(1) does not admit a proof of nonlocality without inequalities. Only when the state in Eq.~(1) happens to be maximally entangled it is possible to exhibit quantum nonlocality without using inequalities, as shown in Refs.~\cite{GHZ,Mermin90bc}.

\end{document}